%% file: conference_101719.tex
\documentclass[conference]{IEEEtran}
\IEEEoverridecommandlockouts
\usepackage{cite}
\usepackage{amsmath,amssymb,amsfonts}
\usepackage{algorithmic}
\usepackage{graphicx}
\usepackage{float}
\usepackage{placeins}
\usepackage{textcomp}

\usepackage{soul}
\usepackage{etoolbox}
\usepackage{array}  

\usepackage{tabularx}
\usepackage{booktabs}
\usepackage{listings}
\usepackage{array}  % For better column formatting

\usepackage{ragged2e} % For text wrapping
\usepackage{url}
\usepackage[table]{xcolor} % Enable table-specific colors
\usepackage{pifont}
\usepackage{booktabs}  % Professional table lines
\usepackage{multirow}  % Merging cells
\usepackage{graphicx,subcaption} 
\usepackage[table]{xcolor}  % Subtle coloring for readability
\usepackage{tikz}
\usepackage{multirow}
\usepackage{enumitem}
\newcommand*\circled[1]{\tikz[baseline=(char.base)]{
            \node[shape=circle,fill,inner sep=2pt] (char) {\textcolor{white}{#1}};}}
\newcommand*\wcircled[1]{\tikz[baseline=(char.base)]{
            \node[shape=circle,draw,inner sep=2pt] (char) {#1};}}
\definecolor{codegreen}{rgb}{0,0.6,0}
\definecolor{codegray}{rgb}{0.5,0.5,0.5}
\definecolor{codepurple}{rgb}{0.58,0,0.82}
\definecolor{backcolour}{rgb}{0.95,0.95,0.95}

\definecolor{lightgray}{gray}{0.92}  % Soft gray for rows
\definecolor{headergray}{gray}{0.80}  % Medium-dark gray for the header
\definecolor{headertext}{gray}{0}  % Pure black text for clarity

\lstdefinestyle{mystyle}{
    backgroundcolor=\color{backcolour},   
    commentstyle=\color{codegreen}\itshape,
    keywordstyle=\color{magenta},
    numberstyle=\color{codegray},
    stringstyle=\color{codepurple},
    basicstyle=\ttfamily\footnotesize,
    breakatwhitespace=true,         
    breaklines=true,                 
    captionpos=b,                    
    keepspaces=true,                 
    numbers=left,                    
    numbersep=4pt,                  
    showspaces=false,                
    showstringspaces=true,
    showtabs=true,                  
    tabsize=2,
    xleftmargin=2em,
    frame=lines,
    framexleftmargin=1.5em,
    basewidth = {.60em},
    moredelim = [s][\color{Blue}]{/**}{*/},
}

\definecolor{lightgray}{gray}{0.9}
\def\BibTeX{{\rm B\kern-.05em{\sc i\kern-.025em b}\kern-.08em
    T\kern-.1667em\lower.7ex\hbox{E}\kern-.125emX}}
\begin{document}

% \title{An Open-Source Benchmark for LLM-aided System Level Design *\\}
% \title{An Open-Source Benchmark for LLM-aided System-on-Chip Design}
\title{SLDB: An End-To-End Heterogeneous System-on-Chip Benchmark Suite \\for LLM-Aided Design}
% \title{SLDB: Can LLMs Design Heterogeneous Systems-on-Chip?}

\author{\IEEEauthorblockN{Elisavet Lydia Alvanaki}
\IEEEauthorblockA{Dept. of Computer Science \\
Columbia University\\
New York, New York, USA \\
ealvanaki@cs.columbia.edu}
\and 
\IEEEauthorblockN{Kevin Lee}
\IEEEauthorblockA{Dept. of Electrical Engineering \\
Columbia University\\
New York, New York, USA \\
kl3266@columbia.edu}
\and 
\IEEEauthorblockN{Luca P. Carloni}
\IEEEauthorblockA{Dept. of Computer Science \\
Columbia University\\
New York, New York, USA \\
luca@cs.columbia.edu}

}
% \and
% \IEEEauthorblockN{2\textsuperscript{nd} Given Name Surname}
% \IEEEauthorblockA{\textit{dept. name of organization (of Aff.)} \\
% \textit{name of organization (of Aff.)}\\
% City, Country \\
% email address or ORCID}

\maketitle

\begin{abstract}
Over the last few years, Large Language Models (LLMs) have emerged as a valuable tool for Electronic Design Automation (EDA). State-of-the-art research in LLM-aided design has demonstrated the ability of LLMs to generate syntactically correct RTL code, showcasing encouraging prospects for integrating AI into the hardware design process. A key enabler of these advancements is the availability of high-quality benchmarks to evaluate new approaches. However, existing datasets and benchmarks fall short of system-level design, as they focus primarily on component-level information and low-complexity designs.
% In these efforts, effective datasets have proved to be of pivotal importance.% 
 To address this gap, we introduce the System-Level Design Benchmark (SLDB), a dataset tailored for evaluating LLMs in system-level integration and configuration tasks. SLDB includes a curated benchmark suite of 10 baseline SoC designs, whose components can be combined into an exponential number of distinct tile-based SoCs through a synthetic library. The dataset provides full SoC configurations, accelerator integration code, communication parameters, and accelerator-aware system configurations, along with testing-application code, compatible with the ESP platform~\cite{ESP}. 
\end{abstract}


\input{sections/00_introduction}
\input{sections/01_related_work}
\input{sections/02_background}
\input{sections/03_methodology}

% \input{sections/04_the_sldb_benchmark}
\input{sections/05_the_sldb_synth_library}

\input{sections/06_experimental}
\input{sections/07_conclusions}
\input{sections/08_appendix}
\bibliography{bibliography} 
\bibliographystyle{ieeetr}
\end{document}

%% file: sections/00_introduction.tex
\section{Introduction}
\label{sec:intro}
Modern hardware systems  are becoming increasingly complex, requiring more efficient design methodologies that balance the trade-offs between performance, energy, and design effort. This complexity has increased with the rise of System-on-Chip (SoC) architectures, which integrate multiple, often heterogeneous components into a single chip to enhance performance across many application workloads. 
While heterogeneous SoCs yield significant gains in performance and efficiency, their design and validation require substantial engineering effort.

Optimized hardware components, designed at the register-transfer level (RTL), demand resource-intensive engineering efforts, characterized by long iteration cycles and extended time-to-market\cite{khailany_dac18}\cite{rashinkar2007system}. These challenges are exacerbated for complex SoCs, underlining the need for more efficient design methodologies.
% EA{ This effort is particularly exacerbated when scaling to Systems-on-Chip. Systems-on-Chip. A System on Chip (SoC) refers to a single-integrated circuit (chip) composed of all the components of an electronic system\cite{definitionSoC}. A SoC is heterogeneous, and typically contains diverse components, such as processor, memory, domain-specific accelerators and more. }. 
In response to this scaling trend, agile design methodologies have emerged, incorporating modularity and design-space exploration techniques to accelerate the design process~\cite{alon20,rautakoura23,agilechipdesign}. However, despite these advances, hardware design remains constrained by manual, human-driven RTL development, limiting automation in the design flow.

Large Language Models (LLMs) have become a promising approach to addressing the growing complexity of hardware design. By leveraging prior designs, LLMs have shown notable results in the generation of RTL designs, making them an increasingly popular research domain in the context of agile hardware design. State-of-the-art works~\cite{blocklove2025automaticallyimprovingllmbasedverilog,10.1145/3649329.3657353,Ho2024VerilogCoderAV,liu2024rtlcodera} have demonstrated the ability of LLMs to generate syntactically and \textit{often} functionally correct code at the component level. As a result, two main challenges have emerged in \textit{LLM-aided design}: 
\\1) Establishing fair and comprehensive evaluation methodologies that assess LLM-generated hardware across diverse design aspects, including correctness, performance, and scalability.
\\2) Expanding LLM-aided design methodologies beyond individual components to complex system-level architectures. %KL - changed "to more complex" to "to complex"

To pave the way for addressing these challenges, we introduce the System-Level Design Benchmark (SLDB). SLDB is a curated dataset that enables the evaluation of LLM performance in system-level design of heterogeneous SoCs. 
In the short term, SLDB provides designers with a valuable resource to assess the performance of LLMs in various system-level design tasks, such as system integration and workload-aware end-to-end system configurations. Designers may use SLDB to evaluate their LLM-aided methodologies post-synthesis (\textit{coarse-grain evaluation}) or to evaluate the ability of these methodologies to generate modular system-level abstractions (\textit{fine-grain evaluation}), such as integration code and system-level design parameters. Designers may also use the SLDB synthetic library to validate more complex applications, dataflows, and component interactions. 
In the longer term, as LLM-aided design progresses, we envision SLDB serving as a solution to benchmark LLM-generated SoC designs against SoCs included in the dataset. SLDB features:
\begin{itemize}[leftmargin=*]
    \item Ten synthesized, heterogeneous SoC designs, integrating diverse workloads, accelerator-aware system configurations, and full-stack communication parameters.
    \item A synthetic library extension that seamlessly combines integrated accelerators to generate an exponential number of distinct SoC designs.
\end{itemize}

%We demonstrate a use of SLDB by completing a comprehensive case study on LLM performance in modular SoC design, using example design tasks from the open-source % %ESP platform~\cite{ESP}. Our analysis provides insights into LLM capabilities for hierarchical and scalable system generation.
Our paper includes a comprehensive case study on LLM performance in modular SoC design using example design tasks from the open-source ESP platform~\cite{ESP}. This study demonstrates how using SLDB provides insights into LLM capabilities for hierarchical and scalable SoC generation.

%% file: sections/01_related_work.tex
\section{Related work}
\label{sec:related}
 Various works have introduced benchmarks and methodologies for evaluating the generation of RTL code. Koios is an advanced benchmark suite for Deep Learning EDA research~\cite{9556371}.  Benchmarks and datasets specifically tailored for LLM-aided hardware design have been developed. In particular, Thakur et al. propose a fine-tuning methodology alongside a testbench-based evaluation framework to assess the functional and syntax correctness of LLM-generated Verilog code of homework-based design questions~\cite{Thakur2022BenchmarkingLL}. VerilogEval is an open-source benchmark using Verilog instructional resources as a data source~\cite{liu2023verilogeval}. Nazzal et al. propose a dataset for systolic array DNN accelerators~\cite{Nazzal2024ADF} that is based on templated generation using Gemmini~\cite{gemmini-dac}. Zhang et al. propose criteria for data selection along with a Verilog dataset with corresponding natural language descriptions~\cite{zhang2024mgverilog}.

Moving towards larger-scale designs, RTLLM is an open-source benchmark for more complex RTL components, along with a prompt engineering methodology to enhance LLM-aided RTL generation~\cite{Lu2023RTLLMAO}.  RTL-Repo is a benchmark using repository-level hierarchy for multi-file RTL projects~\cite{10691810}. 
Met-Rex is a synthesizable benchmark to evaluate reasoning LLMs based on post-synthesis performance metrics~\cite{Abdelatty2024MetRexAB}.
%KL - changed utilizes repository-level to uses

To date, efforts in benchmark development have focused primarily on component-level tasks and lack the architectural complexity necessary to evaluate system-level design methodologies. While suggested \textit{pass rates\footnote{Pass rates (often referred to as $pass@k$) are defined as the percentage of tasks the LLM performs successfully in any of $k$ independent runs (as defined in~\cite{Kulal}). In LLM-aided design, a successful task is defined as a design that is functionally correct.}} for LLM-generated hardware may appear promising, they fail to evaluate designs at the system level, where design parameters, interface specification, timing and communication play a critical role. Current evaluations are limited to self-contained hardware components, whereas realistic, scalable hardware systems functionally depend on the interplay between components, where specification adherence is crucial. \textit{Without a structured system-level benchmark, it remains challenging to assess whether LLM-aided design can effectively scale beyond individual functional blocks to full heterogeneous SoCs.}

%% file: sections/02_background.tex
\section{Background}
\label{sec:background}

\subsection{System-on-Chip Architectures}
Modern SoC architectures feature a growing set of heterogeneous components, including specialized hardware accelerators that offer major performance and energy-efficiency gains~\cite{dally20}. These components impose diverse requirements in terms of area, communication, and delay, and yield distinct design-space tradeoffs. Designers have managed these trade-offs in different manners, aiming to achieve a balance between performance and design effort. In the context of accelerator integration, these tradeoffs have resulted in two dominating paradigms: loosely coupled and tightly coupled accelerators~\cite{7167228}.

% A key aspect of modern SoC design is the integration of hardware accelerators. Modern accelerators can be classified into tightly coupled and loosely coupled architectures, each offering distinct trade-offs in regards to performance, energy efficiency, and engineering design effort. These components impose diverse requirements in terms of area, communication and delay, and yield distinct design space tradeoffs. Designers have managed these tradeoffs in different manners aiming to achieve a balance between performance and design effort. In the context of accelerator integration, these tradeoff has resulted in two dominating paradigms: loosely coupled and tightly coupled accelerators.
\begin{figure}[t]
    \centering
    \includegraphics[width=\columnwidth]{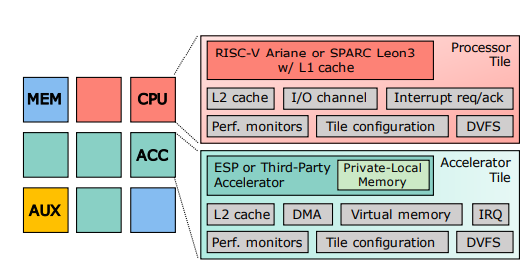}
    \caption{The ESP tile-based architecture \cite{ESP}.}
    \vspace{-2ex}
    \label{fig:esp}
    \vspace{-2ex}
\end{figure}
%Tightly coupled accelerators are directly connected to the processor core, often residing within the same memory hierarchy and are tightly interconnected to the processor, reducing latency. 
Tightly coupled accelerators integrate directly with processor cores and memory hierarchies, significantly reducing latency. However, this tight integration limits flexibility, complicates scaling across workloads, and often requires custom instruction-set architecture (ISA) extensions and compiler modifications, thus increasing engineering effort.

In contrast, \textit{loosely coupled accelerators (LCAs)} operate as independent processing units, typically communicating with the main processor and main memory via high-speed interconnects, such as networks-on-chip (NoC) and dedicated direct memory access (DMA) channels on the NoC. They often perform computation through four main distinct \textit{Accelerator Processing Stages (APS)}: \textit{Configure, Load, Compute and Store}, aimed at reducing communication overhead. During configuration, the accelerator receives the key computation parameters via memory-mapped configuration registers from the software application that has invoked it.
Once the configuration is done, the load stage is initiated and the accelerator sends a DMA request to the memory and receives the requested data via the DMA channel on the NoC. Based on the acquired data, the accelerator performs computations and stores the result in memory via the DMA channel. Some LCAs feature \textit{private local memories (PLMs)}, which are used to reduce NoC traffic. These accelerators are more scalable and modular, allowing SoCs to support a wide range of workloads. 
\begin{table*}[h!]
\centering
\vspace{-3ex}
\renewcommand{\arraystretch}{1.3} % Improve row spacing
\setlength{\tabcolsep}{6pt} % Adjust column spacing for better fit
\begin{tabular}{l m{5cm} r c c c} % Adjust column widths for readability
\toprule
\rowcolor{lightgray} \textbf{Accelerator} & \textbf{Description} & \textbf{Code length} & \textbf{Dataset} & \textbf{Domain} & \textbf{Source Citation} \\
\midrule
AES Encryption & AES encryption algorithm & 13736 & RTL-Repo & Cryptography & \cite{10691810} \\
\rowcolor{lightgray} AES Decryption & AES decryption algorithm & 37658 & RTL-Repo & Cryptography & \cite{10691810} \\
SHA-256 & SHA-256 encryption algorithm & 13343 & RTL-Repo & Cryptography & \cite{10691810} \\
\rowcolor{lightgray} SOBEL & Edge detection filter & 1951 & RTL-Repo & Image Processing & \cite{10691810} \\
FFT & 64-point, 9-stage Fast Fourier Transform & 200911 & RTL-Repo & Image Processing & \cite{10691810} \\
\rowcolor{lightgray} FCDNN & 6-layer DNN with sigmoid activation & 33163 & RTL-Repo & Deep Learning & \cite{10691810} \\
LSTM & Long Short Term Memory layer & 69138 & Koios & Deep Learning & \cite{9556371} \\
\rowcolor{lightgray} SIMPLEDNN & 7-layer DNN with relu activation & 15420 & RTL-Repo & Deep Learning & \cite{10691810} \\
SPMV & MLP network for SPMV & 111074 & Koios & Deep Learning & \cite{9556371} \\
\rowcolor{lightgray} CONVOLUTION & 3-layer convolutional neural network  & 8937 & RTL-Repo & Deep Learning & \cite{10691810} \\
\bottomrule
\end{tabular}
%\caption{Accelerator Specifications}
\caption{The characteristics of the ten accelerators selected to be part of the SLDB benchmark suite.}
\label{tab:accelerator_specs}
\vspace{-3.8ex}
\end{table*}

While LCAs introduce higher communication overhead, they provide greater flexibility in heterogeneous computing, enabling optimized workload distribution and power-efficient execution. 
In developing the SLDB benchmark suite, we focused on modular SoC design based on LCAs and leveraged this modularity to enhance scalability and enable better LLM-design performance.
%KL -- heterogeneous computing environments -- heterogeneous computing
\subsection{A brief overview of ESP}
Due to their flexibility, LCAs have been widely adopted in SoC design. In tandem with these efforts, frameworks have emerged for agile system-level design, allowing designers to build correct-by-construction SoCs by combining many LCAs with processors and memories~\cite{alon20,agilechipdesign}. 
To develop SLDB, we leveraged ESP, an established open-source platform for scalable SoC design and programming~\cite{ESP}. ESP integrates a modular, tile-based architecture with a flexible computer-aided design methodology that includes an RTL-driven design flow. 
%Designers can leverage ESP to generate SoCs by configuring the tiles using the ESP GUI.

The ESP platform allows designers to specify the structure of a target SoC by configuring the number and type of SoC tiles across four main classes: 
CPU tiles, memory tiles, auxiliary tiles (with I/O peripherals such as UART and Ethernet), and accelerator tiles. 
Tile-to-tile communication is performed through a NoC, which supports DMA transfers.

Fig.~\ref{fig:esp} shows an example of the ESP tile-based architecture. ESP promotes design reuse of pre-designed intellectual property (IP) blocks, by seamlessly integrating them into different SoC configurations. For instance, a correctly designed accelerator IP can be integrated in many different ESP SoC design (e.g., SoCs with 2×2, 3×4 or 6x6 tile configurations) together with other accelerators or processing tiles, while remaining compatible with all system components. 
The integration of a heterogeneous component in the tile of a given SoC is performed through a \textit{socket}, which is a parameterized, configurable interface.
A socket interfaces the content of its tile with a multi-plane NoC and provides a variety of platform services~\cite{carloni-dac}.
The gray boxes in Fig.~\ref{fig:esp} illustrate such services—for example, a DMA engine and memory-mapped configuration registers in an accelerator tile socket, and an L2 cache and interrupt controller in a CPU tile socket.

While ESP facilitates the agile development of SoCs, most application-dependent design tasks must be performed manually.
In particular, the SoC integration of a given accelerator requires the completion of two manual tasks starting from two provided templates: 
(1) the design of an RTL wrapper that interfaces the accelerator logic with the socket logic (including the DMA engine and the NoC interface)
and (2) the implementation of a device driver to expose accelerator-specific parameters to software applications.
The automatic completion of both these tasks is the target of the LLM-aided design techniques discussed in Section~\ref{sec:sldb}.

%% file: sections/03_methodology.tex
\section{The SLDB Benchmark Suite}
\label{sec:sldb}

\begin{figure*}[t]
    % \centering
    \vspace{-2ex}
    \includegraphics[width=1\textwidth]{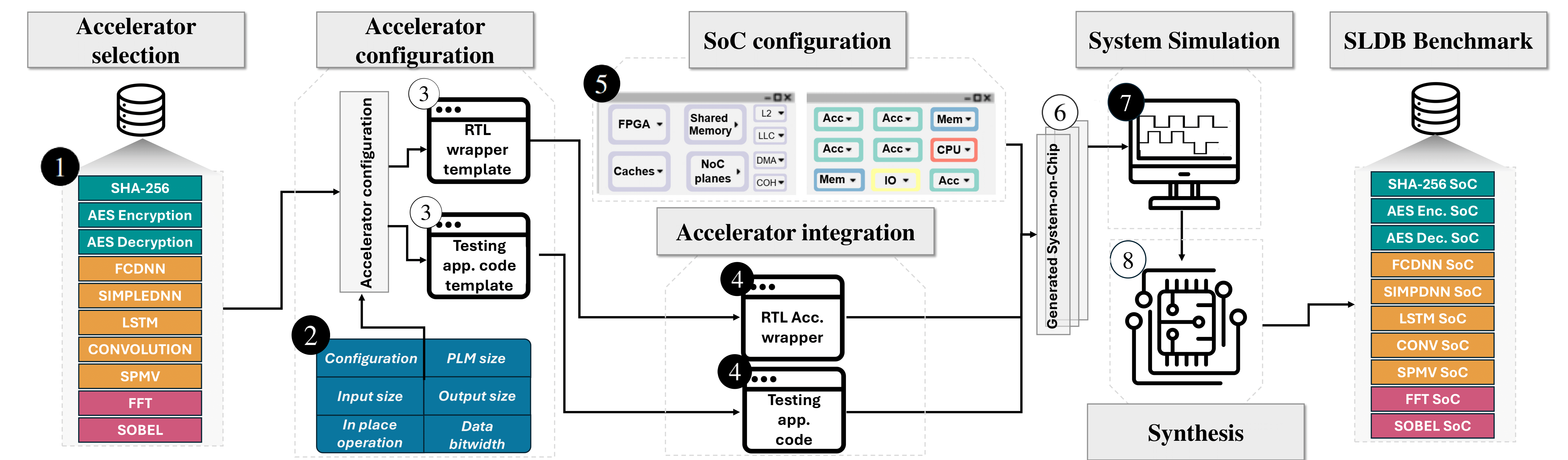}
    \caption{SLDB design flow. Steps marked with a black-filled circle require manual configuration. Steps marked with a white-filled circle are generated by ESP.}
    \vspace{-2ex}
    \label{fig:esp_flow}
    \vspace{-2ex}
\end{figure*}

\subsection{Accelerator Data Acquisition}
The SLDB benchmark suite includes complex accelerator designs from various domains, such as deep learning, cryptography and image processing. 
We selected a curated set of accelerators from the RTL-Repo and Koios benchmarks, to use as accelerators to be integrated into an SoC. 
We filtered out programmable accelerators and accelerators that are not suitable for LCA integration. 
Our goal is to evaluate LLM performance on system-level integration as well as system configuration tasks. 
We opted to select some long-context, complex accelerators from the Koios benchmark, to assess the ability of the LLM to cater to designs with larger contexts. SLDB aims to evaluate LLM performance in realistic scenarios, with accelerators running complex and resource-intensive workloads. The selected designs illustrate diversity in terms of communication requirements, as well as workload properties. The structure of the dataset is illustrated in Table~\ref{tab:accelerator_specs}.

We selected designs based on the following criteria:

\textbf{Suitability for Loosely-Coupled Acceleration:}
Accelerators must be compatible with loosely-coupled integration, without direct dependencies on the processor's ISA or tightly coupled memory hierarchies. This requirement ensures compatibility with ESP.

\textbf{Workload Complexity:}
Accelerators reflect computationally intensive workloads, as used in modern SoCs. These workloads can effectively test integration and include diverse system configuration requirements.

\textbf{Completeness, Quality, and Open-Source Availability:}
Selected accelerators must be complete, freely available, and adhere to good coding practices. Designs with incomplete source code, closed-source dependencies, or low-quality codebases were excluded to maintain the benchmark’s usability.

\subsection{Integration into an ESP SoC}
We structured the integration process into eight distinct stages, as shown in Fig.~\ref{fig:esp_flow}. 
Steps that require manual integration are denoted by circles with a fill, and automated steps are denoted by no fill. In Step \circled{1}, the accelerators are selected from the RTL-Repo and Koios benchmarks. In Step \circled{2}, the accelerator configuration is defined. In Step \wcircled{3}, the templates for the C testing application code (baremetal and Linux) are generated, along with a Verilog template for the integration of the RTL accelerator. The configurations are also written into the ESP files, to be used to generate the SoC in subsequent stages. In Step \circled{4}, the accelerator wrapper template is modified to integrate the accelerator. This includes the generation of RTL code to manage DMA handshakes, establishing mappings between the accelerator's individual load, compute, and store operations and the packed data transfers performed via the DMA bus, as well as implementing the necessary timing and internal synchronization logic to align accelerator operations with the Configure-Load-Compute-Store model. The testing application code is optionally modified to initialize the input and golden output values. In the context of this benchmark, we use randomized values to test the accelerators, since their functionality has already been verified. In Step \circled{5}, the heterogeneous SoC configuration is selected through the ESP GUI. The components are selected through drop-down options, with no code modifications required. In Step \wcircled{6} the full SoC RTL is generated. In Step \circled{7} the system is simulated and debugged to ensure functionality. Finally, in Step \wcircled{8} the designs are  synthesized for FPGA to retrieve performance, power and area (PPA) metrics.
\begin{table}[t]
    \centering
    \renewcommand{\arraystretch}{1.2}
    \resizebox{\columnwidth}{!}{%
    \begin{tabular}{m{1.5cm} c c c}
        \toprule
        \rowcolor{lightgray}

        {\centering\textbf{SoC}} &
        {\centering\textbf{Utilization (LUTs)}} &
        {\centering\textbf{Static Power}} &
        \multicolumn{1}{m{2.5cm}}{\centering\textbf{Max Datapath Delay}} 
         \\
        \midrule
        \rowcolor{lightgray}{\centering AES Enc.} & {\centering 112,008} & {\centering 4.3 W} & {\centering 32.94 ns} \\
        {\centering AES Dec.} & {\centering 120,539} & {\centering 5.83 W} & {\centering 72.65 ns} \\
        \rowcolor{lightgray}{\centering SHA-256} & {\centering 99,699} & {\centering 2.87 W} & {\centering 10.01 ns} \\
        {\centering SOBEL} & {\centering 98,071} & {\centering 2.86 W} & {\centering 19.26 ns} \\
        \rowcolor{lightgray}{\centering FFT} & {\centering 109,627} & {\centering 3.39 W} & {\centering 3.8 ns} \\
        {\centering FCDNN} & {\centering 105,618} & {\centering 2.93 W} & {\centering 19.8 ns} \\
        \rowcolor{lightgray}{\centering LSTM} & {\centering 149,911} & {\centering 3.96 W} & {\centering 9.63 ns} \\
        {\centering SIMPLEDNN} & {\centering 98,518} & {\centering 2.88 W} & {\centering 9.59 ns} \\
        \rowcolor{lightgray}{\centering SPMV} & {\centering 103,940} & {\centering 2.94 W} & {\centering 32.94 ns} \\
        {\centering CONVOLUTION} & {\centering 120,311} & {\centering 2.9 W} & {\centering 19.37 ns} \\
        \bottomrule
    \end{tabular}
    }
    \caption{Performance and utilization of the ten SLDB SoCs.}
    \vspace{-1ex}
    \label{tab:soc_performance}
    \vspace{-3ex}
\end{table}

\subsection{The SLDB Benchmark Structure}
\label{sec:sldb_baseline}
The SLDB benchmark suite consists of ten heterogeneous SoCs.
Each SoC has a \textit{baseline} ESP 2x2 tile-based architecture, with four tiles: a CPU tile with the Ariane RISC-V core~\cite{zaruba19}, a memory tile, an auxiliary tile and the accelerator tile, which contains one curated LCA from Table~\ref{tab:accelerator_specs}.
The SoC designs have been first evaluated in simulation, then synthesized and implemented with Vivado 2023.2 to retrieve PPA metrics. 
Table~\ref{tab:soc_performance} reports post-implementation metrics.

We evaluated the correctness of the interfacing and SoC parameters, as well as the testing application code, but assumed correctness at the accelerator level, as the accelerators have been retrieved from pre-existing peer-reviewed works~\cite{9556371,10691810}. 
Each design includes the full integration into an SoC using the design flow outlined in Fig.~\ref{fig:esp_flow}. The selected and available configurations for the SoCs are reported in Table~\ref{tab:system_config}. The SoCs can run in bare-metal mode or with the Linux operating system, allowing designers to evaluate design methodologies that scale up to running software applications on top of the Linux operating system. %level. 

\begin{figure}[t]
    \centering
    \vspace{-3ex}
    \includegraphics[width=0.9\columnwidth]{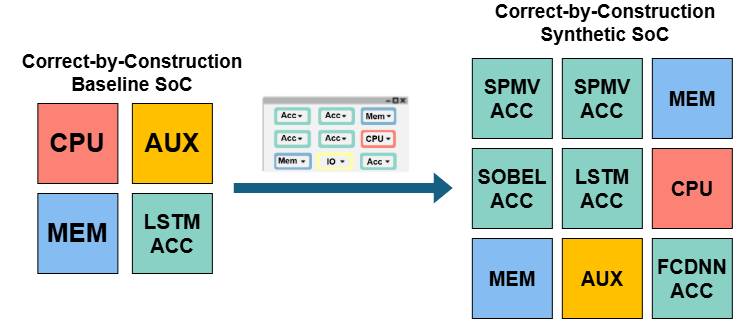}
    \caption{SLDB synthetic library functionality.}
    \vspace{-2ex}
    \label{fig:esp_3x3}
    \vspace{-2ex}
\end{figure}

% \subsection{Considerations of Using SLDB\EA{remove?}}
% With the benchmark structure in place, we now describe our approach for leveraging SLDB to assess LLM-aided methodologies. In Section~\ref{sec:intro}, we introduced two evaluation strategies: The first is a \textit{coarse-grain evaluation}, which focuses on post-synthesis analysis. The second is a \textit{fine-grain evaluation}, which examines the methodologies’ ability to generate modular system-level abstractions. In addition, designers can leverage SLDB synthetic library for validation. More specifically, it can be used to compare LLM-aided design implementation in the context of a full SoC, and monitor parameters such as NoC congestion, LCA memory access penalties, as well as other system parameters. In Section~\ref{sec:experimental}, we perform a case-study based on a fine-grained evaluation using ESP.
% \subsection{Comparison to existing RTL benchmarks}
% Existing bechmarks such as, 
% aim to evaluate LLMs on elementary 
% hardware design tasks, and often 
% include few and isolated components, 
% where integration and system 
% interaction is not a critical issue. 
% In current works that focus on 
% larger codebases, the components are not integrated in a system context, where communication protocol adherence, timing and configuration becomes increasingly relevant. 
% \cite{Lu2023RTLLMAO}.

%% file: sections/05_the_sldb_synth_library.tex
\begin{figure*}[t]
    % \centering
    \vspace{-2ex}
    \includegraphics[width=1\textwidth]{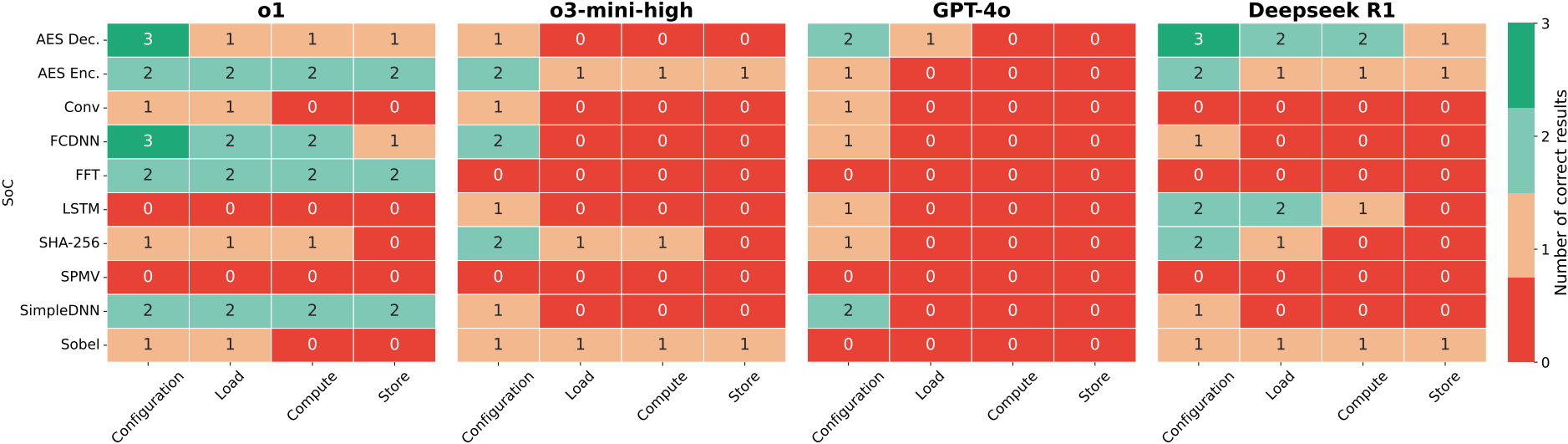}
    \caption{Distribution of errors per Accelerator Processing Stage for 3 LLM generation runs.}
    \vspace{-2ex}
    \label{fig:heatmap-stages}
    \vspace{-1.5ex}
\end{figure*}

\subsection{Scaling to different SoCs - The SLDB Synthetic Library}
 % Designs can be modified through the ESP GUI to alter configurations as outlined in Table~\ref{tab:system_config}, select different target hardware and more, as outlined in Section~\ref{sec:sldb_synth_lib}.
SLDB users may start from the baseline SoC configurations to create many different SoCs through the ESP GUI, as shown in Figure~\ref{fig:esp_3x3}. 
Any combination of the baseline accelerators may be used with any amount or type of the ESP-compatible CPUs, memory, and auxiliary tiles. 
The available configurations are listed in Table~\ref{tab:system_config}. 
Any configurations listed in the table can be selected along with the integrated accelerators without modifications to the existing code. The selection can be made by repeating Step \circled{3} of the SLDB design flow and resynthesizing the design. Due to the correct-by-construction design capabilities of ESP, any SoC combination offers guaranteed correctness assuming a correctly integrated baseline accelerator and correct accelerator-specific configuration (provided in the benchmark).

%% file: sections/06_experimental.tex
\begin{table}[h]
\centering
\renewcommand{\arraystretch}{1} % Improve row spacing
\setlength{\tabcolsep}{1pt} % Adjust column spacing for better fit
\begin{tabular}{l m{4cm}} % Adjusting column width for better readability
\toprule
\rowcolor{lightgray} \textbf{Configuration} & \textbf{Selected / Range} \\
\midrule
 FPGA Prototyping & 
xilinx-vc707 \newline (profpga-xc7v2000t, 
xilinx-zcu102, xilinx-
zcu106, profpga-xcvu19p, profpga-
xcvu440, xilinx-vc707t,  xilinx-vcu118, 
xilinx-vcu128) \\
\rowcolor{lightgray} CPU & Ariane \newline (Ibex, SPARC) \\
Shared Local Memory (per tile) & 256 KB \newline (64-4096 KB, powers of 2) \\
\rowcolor{lightgray} Accelerator Data Allocation Strategy & Scatter/Gather \newline (Big Physical Area) \\
Caches & No \newline (LLC, L2, Accelerator L2 configurable) \\
\rowcolor{lightgray} SoC Dimensions (rows × columns) & 2 × 2 \newline (up to 8 × 8) \\
 Coherence NoC Planes Bitwidth & 64 \newline (32-1024, powers of 2) \\
\rowcolor{lightgray} DMA NoC Planes Bitwidth & 64 \newline (32-1024, powers of 2) \\
 Multicast & - \newline (up to 16) \\
\bottomrule
\end{tabular}
\vspace{-1ex}
\caption{Selected system configuration parameters. Available options are illustrated in parentheses.}
\label{tab:system_config}
\vspace{-3ex}
\end{table}

\section{Case study - SoC Design with \\Task Decomposition}
\label{sec:experimental}

We adopt a task decomposition~\cite{10.5555/3600270.3602070} strategy and propose a \textit{fine-grain} LLM evaluation approach structured around the ESP platform. This allows us to effectively assess the current capabilities of LLMs across diverse, modular hardware design tasks, while also enabling performance improvements through decomposition. Due to the complexity of end-to-end SoC design, we do not evaluate out-of-the-box LLMs in the full design task, as we consider current progress
in the domain to be inadequate for successful completion of the task without additional optimizations.

To explore the capabilities of current LLMs in SoC design aspects, we use SLDB to perform a case study of their functional and syntax correctness in a system-level design task. The LLM models are tasked with completing the integration of accelerators into ESP and evaluating required system parameters given the baseline accelerators. This case study evaluates:
\begin{itemize}[leftmargin=*]
    \item The ability of LLMs to abstract system-level parameters from RTL accelerators.
    \item The performance of LLMs in integration tasks with standardized protocols (such as DMA).
    \item The LLMs' ability to evaluate correct signal-port mappings and facilitate data transfers to and from the accelerator.
\end{itemize}

% However, it should be noted that the benchmark suite is \textit{orthogonal to the ESP methodology}, and may be used for performance, area, and correctness validation of SoC generated with other methodologies, modular or otherwise. \EA{we need to explicitly state and emphasize that this benchmark can be used for any methodology, we evaluate using ESP to make the task more decompositioned}

\subsection{Evaluation Metrics}
Functional correctness for each accelerator integration task is evaluated through the four sequential APSs defined in Section~\ref{sec:background}.
Each APS represents a critical step in ensuring accurate communication and integration of accelerators within the ESP framework. The APSs evaluations are defined as follows:
\begin{enumerate}[leftmargin=*]
    \item \textit{Configuration}: This stage evaluates whether the LLM-generated integration correctly assigns configuration parameters and initializes accelerator-specific registers required for correct operation.    
    \item \textit{Load}: The correctness of data loading through DMA is evaluated by verifying if the accelerator correctly issues DMA requests, adheres to the appropriate handshake protocols, and receives the intended input data from memory.
    \item \textit{Compute}: This stage evaluates whether the integration code correctly maps accelerator input/output ports, accurately manages synchronization signals, and ensures proper computation within the accelerator logic.
    \item \textit{Store}: This stage assesses if the accelerator successfully returns computed results back to memory via DMA, correctly handling outgoing data and ensuring synchronization during transfers.
\end{enumerate}

An integration is marked as fully functionally correct if all four APSs complete successfully. Correctness is assessed through simulation, and validated against integration by a human engineer. If a given stage fails, subsequent dependent stages are noted as failed due to dependency.

\subsection{LLM Evaluation}
We perform experiments to evaluate LLM performance using SLDB. The evaluated LLMs include a mixture of closed and open-sourced models, as well as reasoning and non-reasoning models. We evaluate OpenAI o1, GPT-4o, o3-mini-high and DeepSeek-R1-671B.The prompt used is identical across inferences and models. Due to space limitations, the full prompt will be included in the open-source release. Each LLM is tasked with generating DMA integration wrappers, extracting accelerator-specific configuration parameters, and generating driver code, for all ten SLDB accelerators. We conduct three independent inference runs per task, per accelerator, per LLM model and report pass rates (pass@1 and pass@3). We use the benchmark designs as our baseline.

Table~\ref{tab:model-performance-overall} compares the evaluated LLM models, presenting pass@1 and pass@3 metrics for syntactic and functional correctness. 
While syntactic correctness remains high, functional correctness varies across models and poses a significant challenge, with the highest pass rate at 50\%. The pass@3 metric significantly improves performance for some models, underlining the importance of multiple runs in fair evaluation of model performance.

\begin{table}
\centering
\begin{tabular}{lcccc}
\hline
\rowcolor{lightgray}\textbf{Model} & \multicolumn{2}{c}{\textbf{pass@1}} & \multicolumn{2}{c}{\textbf{pass@3}} \\ 
\rowcolor{lightgray} & \textbf{Syntax} & \textbf{Functional} & \textbf{Syntax} & \textbf{Functional} \\ 
\hline
o1 & 70\% & \textbf{40\%} & 90\% & \textbf{50\%} \\
\rowcolor{lightgray}o3-mini-high & \textbf{100}\% & 10\% & \textbf{100\%} & 20\% \\
GPT-4o & 60\% & 0\% & 70\% & 0\% \\
\rowcolor{lightgray}DeepSeek-R1-671B & 60\% & 0\% & 60\% & 30\% \\
\hline
\end{tabular}
\caption{Performance comparison of models for pass@1 and pass@3 }
\vspace{-1ex}
\label{tab:model-performance-overall}
\end{table}

\begin{figure}[t]
    \centering
    \includegraphics[width=\linewidth]{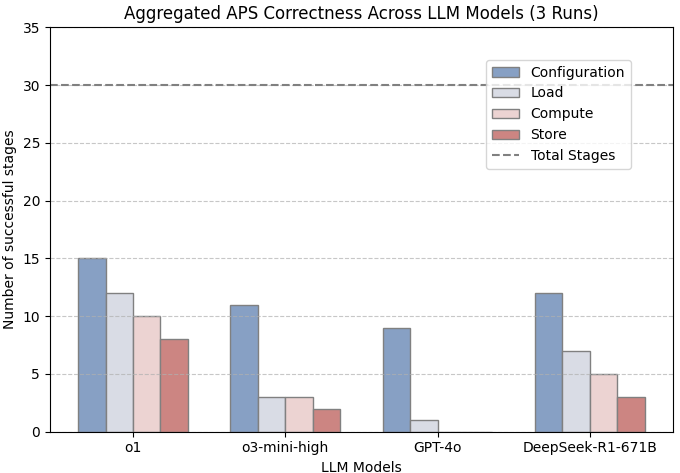}
    \caption{Number of SoC integration errors per model. Experiments conducted using 3 independent inference runs.}
    \vspace{-1ex}
    \label{fig:enter-dist-dma-stage}
    \vspace{-3ex}
\end{figure}

 APSs successfully completed over three inference runs are shown in Figures~\ref{fig:heatmap-stages}~and~\ref{fig:enter-dist-dma-stage}. 
Fig.~\ref{fig:heatmap-stages} shows successful APSs for each SoC, per model. A value of 3 means that all generation runs were successful. We observe that o1 outperforms all models, yet still fails on more challenging accelerators, such as LSTM and SPMV. 
Fig.~\ref{fig:enter-dist-dma-stage} shows the total number of successful APSs per model. It is evident that o1 and Deepseek R1 showcase minimal performance degradation across APSs, while other models show significant performance degradation in the load APS, where the DMA handshake begins. 
For Figures~\ref{fig:heatmap-stages} and~\ref{fig:enter-dist-dma-stage}, the designs that complete the store APS successfully are also functional. \textit{These results are cumulative over all design runs, whereas the pass@3 results consider non-overlapping successful design results across different LLM generation runs}.

\begin{figure}[h!]
    \centering
    \vspace{-2ex}
    \includegraphics[width=\linewidth]{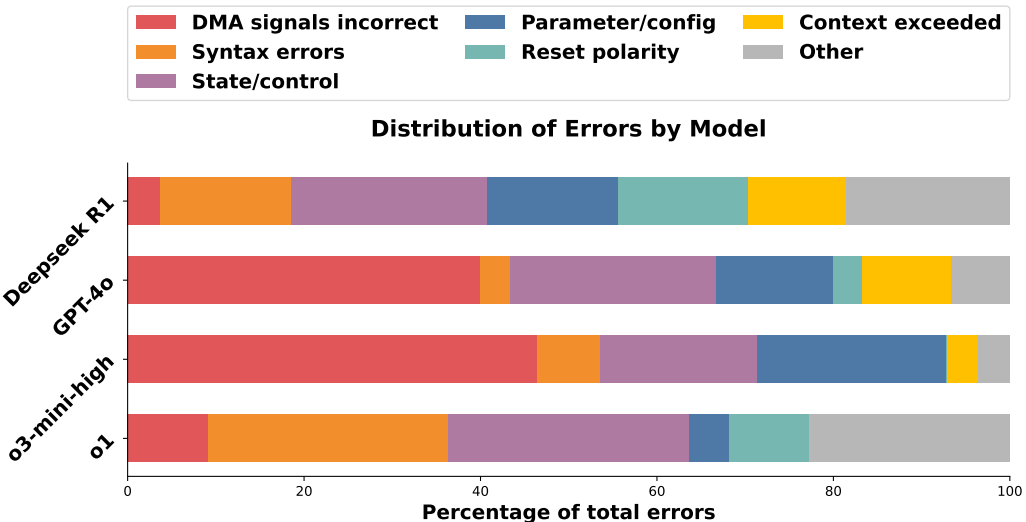}
%    \caption{Distribution of errors per model. Experiments conducted using 3 independent generation runs.}
    \caption{Distribution of errors per model.}
    \vspace{-1ex}
    \label{fig:dist-error-type}
    \vspace{-1ex}
\end{figure}

Fig.~\ref{fig:dist-error-type} categorizes the types of errors encountered for each LLM across all runs. Prominent error types include incorrect handling of DMA signals, DMA state transitions, and specification adherence issues. 
Notably, advanced models like o1 and DeepSeek-R1 present fewer DMA signal errors, while most models struggle significantly with the complexity of state and control logic. Finally, specification adherence, such as active-low reset polarity, seems to become challenging for models as the length of the design code increases.

\begin{figure}[h]
    \centering
    \includegraphics[width=0.8\linewidth]{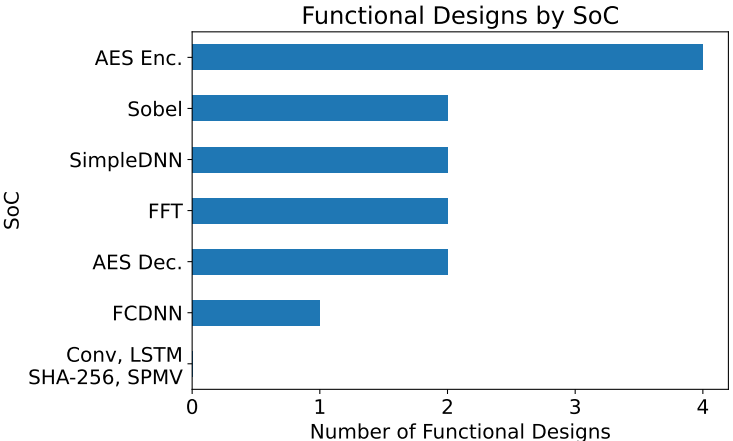}
    \caption{Number of correct designs per SoC type}
    \vspace{-2ex}
    \label{fig:correct_per_acc}
    \vspace{-2ex}
\end{figure}

As shown in Fig.~\ref{fig:enter-dist-dma-stage}, while most models perform well in the configuration stage, their performance degrades in transition-defined APSs, such as the DMA handshake. The steep drop in success rates highlights the challenges LLMs face when handling complex DMA handshakes and state transitions, suggesting areas for targeted improvements. 
Fig.~\ref{fig:correct_per_acc} shows correctly generated accelerators, counted cumulatively for all models, across all runs. Results clearly indicate that certain accelerators, especially those with complex internal logic or advanced synchronization requirements (e.g., LSTM, SPMV, SHA-256), pose challenges to LLMs, underscoring areas for targeted improvement in future work.

%% file: sections/07_conclusions.tex
\section{Conclusions and Future Work}
\label{sec:conclusionsandfuture}
SLDB establishes a structured benchmark suite for evaluating LLM-aided SoC design at the system level. Unlike existing benchmarks, SLDB enables a comprehensive assessment of LLM performance in realistic and diverse system design tasks, including heterogeneous SoC configuration, accelerator integration, and system-level communication.

To demonstrate its utility, we conducted a case study on LLM-based fine-grain accelerator integration. Although LLMs achieved high syntax correctness, they struggled with system-level constraints, particularly DMA transactions, state transitions, and system-level parameter extraction. These challenges emphasize the need for specification-aware techniques such as streaming LLMs \cite{xiao2023streamingllm} to improve long-context understanding.

SLDB serves as a benchmarking tool for evaluating the performance of LLM-aided design methodologies and guiding future improvements in AI-driven hardware design. Future work in this direction includes expanding SLDB for broader system-level evaluations. SLDB is open-sourced\footnote{\url{https://github.com/sld-columbia/SLDB}} and will be maintained and improved as an open-source project to support continued research in LLM-aided system integration.

%% file: sections/08_appendix.tex
\pagebreak

\section{Appendix}
\label{sec:appendix}

\subsection{In-Context Learning Evaluation}
We perform an evaluation of the integration capabilities of LLMs through In-Context Learning (ICL), a technique that enables LLMs to leverage domain-specific context to improve performance on a given task~\cite{NEURIPS2020_1457c0d6}. We perform In-Context-Learning by providing the models with correct wrappers from the SLDB, and prompting them to complete the generation of template wrappers for the remaining designs. The following prompt is used for In-Context Learning: \textit{"I will provide you with 3 reference accelerator wrappers, integrating 3 different accelerators into a system through DMA. Based on these wrappers, modify the input wrapper template that I will give you to integrate the input accelerator that I will give you. OUTPUT THE FINAL CODE AS IT WILL BE COMPILED AND TESTED. RESET FOR THE WRAPPER IS ACTIVE LOW."}

We use three accelerator wrappers (for the FCDNN, LSTM and SOBEL) as context, and evaluate the integration performance for the remaining seven accelerators: 
each accelerator is integrated in the SLDB baseline 2x2 tile SoC configuration described in Section~\ref{sec:sldb_baseline}.
Table~\ref{tab:icl_performance} reports the results of the ICL evaluation.

\begin{table}[h]
    \centering
    \begin{tabular}{lcc}
        \toprule
        Model & Functional correctness & Syntax correctness\\
        \midrule
        o1 & 0.00\% & 100\% \\
        o3-mini-high & 14.20\% & 71\% \\
        GPT-4o & 0\% & 71\% \\
        Deepseek-R1 & 0\% & 43\% \\
        \bottomrule
    \end{tabular}
    \caption{Comparison of Model Performance with ICL}
    \label{tab:icl_performance}
\end{table}

\begin{figure}[t]
    \centering
    \includegraphics[width=\linewidth]{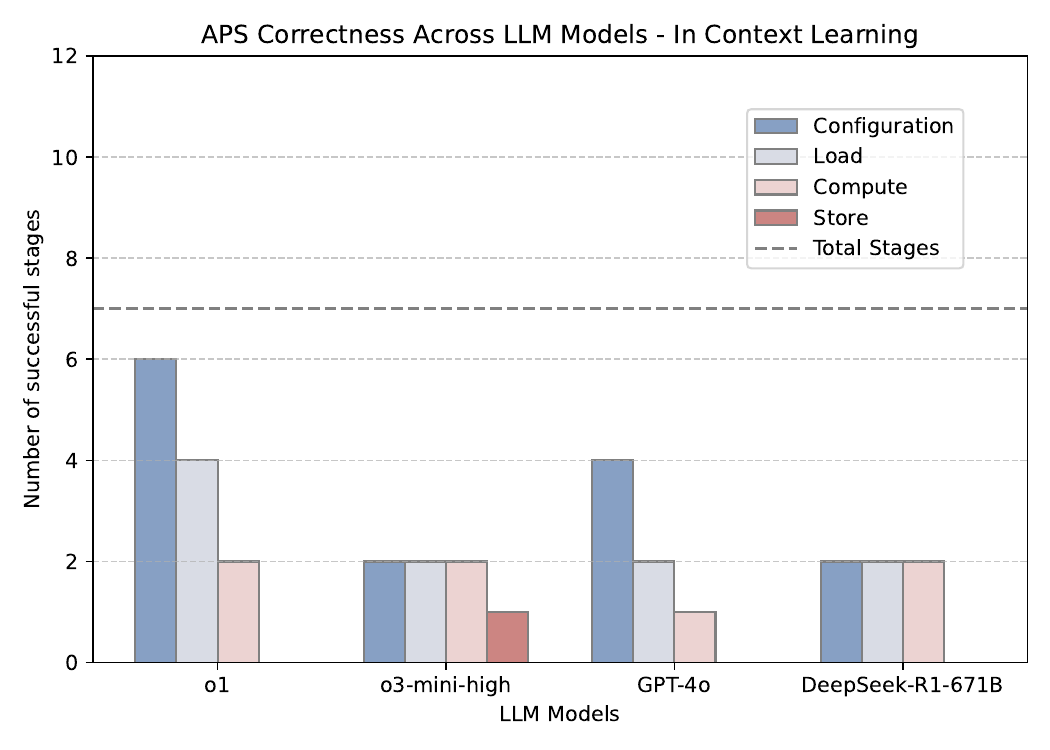}
    \caption{Summary of correctly generated DMA stages, for 7 SoCs with In Context Learning.}
    \label{fig:icl-dma-bar}
\end{figure}

Figure~\ref{fig:icl-dma-bar} shows the distribution of the correctly generated APSs per model. The horizontal dashed line denotes the total number of SoCs, for which the LLM is tasked with completing the APSs. 
If a processing APS fails, then the subsequent APSs are classified as ''failed'' too, due to the strict functional dependencies among the four APSs.
Figure~\ref{fig:icl-dma-bar} highlights each LLM model's capability to successfully integrate the four APSs using in-context learning -- Configuration, Load, Compute, and Store. o1 exhibits the best progress, correctly completing six, four, and two APSs for configuration, load, and compute, respectively. GPT-4o displays a similar trend, but with lower overall performance. o3-mini-high and DeepSeek-R1-671B exhibit less completed APSs overall, each successfully completing only two APSs for the configuration, load, and compute APS. Notably, the o3-mini-high model is the only model that successfully integrated an accelerator from the configuration APS throughout the store APS. 
This analysis is a snapshot on the current level of performance exhibited by LLMs when applied to the task of integrating an LCA into an SoC using in-context learning. While o1 is the best performing LLM among the four evaluated models, it is clear that more progress remains to be made in this interesting avenue of research for the LLM-aided design community.

